# Unified Performance Analysis of Hybrid FSO/RF System with Diversity Combining

Long Huang[1,2], *Member, IEEE*, Siming Liu[3], Pan Dai[1,2], Mi Li[1,2], Gee-Kung Chang[4], Fellow, IEEE, Yuechun Shi[1,2], Xiangfei Chen[1,2], *Senior Member, IEEE*

[1]*Key Laboratory of Intelligent Optical Sensing and Manipulation, Ministry of Education, Nanjing 210093, China*
[2]*School of Engineering and Applied Sciences, Nanjing University, Nanjing 210093, China (e-mail: shiyc@nju.edu.cn)*
[3]*Beijing University of Posts and Telecommunications, Beijing 100876, China*
[4]*School of Electrical and Computer Engineering, Georgia Institute of Technology, Atlanta, GA 30308, USA*
[*]*shiyc@nju.edu.cn*

**Abstract:** Hybrid free space optical (FSO)/radio frequency (RF) systems have been proved to be reliable links for high-data-rate wireless backhauls. In this paper, we present a unified performance analysis of the hybrid FSO/RF transmission system which transmits the identical data in both links and implements two popular diversity combining schemes, namely, selection combining (SC) and maximal ratio combining (MRC), in the receiver. Specially, for the FSO link, the Gamma-Gamma turbulence with pointing errors under heterodyne detection (HD) and intensity modulation/direction detection (IM/DD) is considered in our analysis while the general κ-μ shadowed fading which unifies popular RF fading models is employed for the analysis of the RF link. As a result, unified closed-form expressions of outage probabilities and average bit error rates for different modulation schemes are derived. Analytical and Monte Carlo simulation results are provided to characterize the performance of the hybrid FSO/RF link which is compared to the single FSO link and the single RF link. The agreement between the analytical and simulation results confirms the unification of various FSO channels and RF fading scenarios into a single closed-form expression.

**OCIS codes:**

## References and links

## 1. Introduction

In recent years, free space optical (FSO) communications have gained significant importance owing to the unique features: large bandwidth, ease of deployment, license free spectrum, lower power, less mass, small size, and improved channel security [1]. Despite these desirable advantages, FSO links are greatly affected by fading due to atmospheric turbulence and pointing errors [2, 3]. Turbulence-induced fading, known as scintillation, causes irradiance fluctuations in the received optical signal as a result of variations in the atmospheric refractive index [2]. Furthermore, dynamic wind loads and weak earthquakes cause vibrations of the optical beam, leading to random irradiance fluctuations in the received optical signal [3]. By employing multiple FSO links, the performance of FSO communication systems can be improved [4, 5]. However, all the FSO links still suffer from fog weather. A possible solution to improve the FSO link's reliability is to integrate it with a millimeter-wave (MMW) radio frequency (RF) link since the MMW RF link exhibits complementary characteristics to weather conditions and supports data rates similar to those of FSO. Specifically, the FSO link degrades significantly due to fog but is not sensitive to rain while the MMW RF link is very sensitive to rain but is quite indifferent to fog [6]. As a result, FSO and RF transmission systems are good candidates for joint deployment to provide reliable high data rate communications.

For the hybrid FSO/RF architecture, several studies have been reported on the system design. In [8-10], the data are jointly coded and divided into two streams which are transmitted by the FSO link and the RF link, respectively. The soft-switching between the two links has a significant improvement in the total link capacity. However, it is not suitable to implement the soft-switching in the fiber-wireless integrated mobile backhaul (MBH) architecture proposed in [7]. In the fiber-wireless integrated MBH architecture, the hybrid FSO/RF link provides a reliable link between the core network and base stations (BSs) where the RF signal is generated by photonic methods, leading to a scalable and cost-effective network setup. Therefore, if the soft-switching is employed in the fiber-wireless integrated MBH, the unencoded data carried by the optical carrier should be delivered from the core network to BSs. In the BS, the optical signal is converted to an electrical signal, and then a joint coding scheme is implemented, leading to two electrical coded streams. One coded stream is up-converted to the MMW band by an RF front-end for the RF link while the other stream is converted back to an optical signal by an optical front-end for the FSO link. In the network architecture proposed in [7], the optical data stream is delivered to the BS from the core network and directly emitted for the FSO link, and the optical data stream can be up-converted to the MMW RF signal at the same time by the photonic method, resulting in a simplified front-end design [11]. Based on this consideration, the scenario of transmitting identical data in both links is more suitable in the fiber-wireless integrated MBH. Previously, hard-switching between FSO and RF links which transmit the same data is proposed in [12]. In this scheme, the FSO link is used as the primary transmission channel as long as its signal-to-noise (SNR) is above a certain threshold value. When the SNR of the FSO link falls below the threshold, the receiver sends a feedback signal to activate the RF link for data transmission. To avoid the feedback transmission stage, diversity combining of the FSO link and the RF link can be implemented in the receiver. Two popular diversity combining schemes, namely selection combining (SC) and maximal ratio combining (MRC), are analyzed for the hybrid FSO/RF link with phase shift keying (PSK) modulation [13]. An outage analysis of the hybrid FSO/RF link is presented with MRC in [14]. The SC is also evaluated in the hybrid FSO/RF system [15]. In terms of the FSO channel, pointing errors and two detection types, namely heterodyne detection (HD) or intensity modulation/direct detection (IM/DD), are not considered in [13-15]. As to the RF channel, the Rician fading is considered in [13, 14], and the Rayleigh fading is considered in [15]. The MRC receiver used to combine the hybrid FSO/RF link is experimentally demonstrated in [7]. Although the atmospheric turbulence fading is emulated for the FSO link in [7], the pointing error, the detection type, and the RF fading are not considered in the experiment.

In this paper, we present a detailed outage and error rate analysis of the hybrid FSO/RF system where SC or MRC is used to combine the FSO and RF sub-links. For the FSO sub-link, the Gamma-Gamma turbulence-induced fading with pointing errors under HD and IM/DD is considered in our study. Moreover, a very general RF fading model which unifies many popular fading models including one-side Gaussian, Rayleigh, Nakagami-$m$, Rician, $\kappa$-$\mu$, and Rician shadowed fading is used to analyze the hybrid FSO/RF link [16, 17]. To the best of our knowledge, this is the most general hybrid FSO/RF channel considered in the literature which can include the previous investigations as special cases [13-15]. Novel closed-form results for the outage probability (OP) and average bit error rates (BERs) for different modulation schemes are derived in terms of the Fox's H function. Using the derived formulas, the obtained numerical results show the hybrid FSO/RF link achieves better reliability and performance than the single FSO link and the single RF link.

## 2. Principle

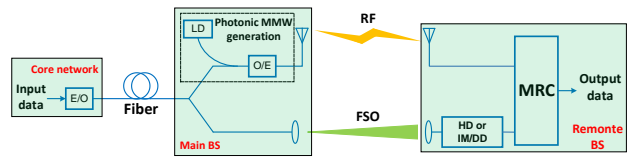

**Fig. 1. Schematic diagram of the hybrid FSO/RF link incorporated in the fiber-wireless integrated MBH. LD: laser diode, E/O: electrical-to-optical conversion, O/E: optical-to-electrical conversion.**

The schematic diagram of the hybrid FSO/RF system integrated in the fiber-wireless MBH proposed in [7] is shown in Fig. 1. The input data are delivered by the optical fiber from the service gateway and mobility management entity (S-GW/MME) of the core network to the main BS where the optical signal can be directly emitted by an optical aperture while an MMW signal can be generated by the photonic method as shown in Fig. 1 [7, 11]. The photonic generation of MMW signals is widely used in high speed fiber-wireless systems [11]. As a result, a hybrid FSO/RF link is established between the main BS and the remote BS to achieve reliable MBH.

*A. FSO Subsystem*

Firstly, the FSO sub-system is investigated. The FSO link is assumed to be subject to Gamma-Gamma fading that accounts for pointing errors and both types of detection techniques (i.e. HD as well as IM/DD). Therefore, the probability density function (PDF) of the receiver irradiance $I$ can be given by [3, Eq. (1)]

$$f_I(I) = \frac{\xi^2}{I\Gamma(\alpha)\Gamma(\beta)} G_{1,3}^{3,0}\left[\alpha\beta \frac{I}{A_0} \middle| \begin{matrix} \xi^2+1 \\ \xi^2, \alpha, \beta \end{matrix} \right], \tag{1}$$

where $\xi$ is the ratio between the equivalent beam radius and the pointing error displacement standard deviation at the receiver, $A_0$ is a constant term that represents the pointing loss, $\alpha$ and $\beta$ are the fading/scintillation parameters related to the atmospheric turbulence conditions with lower values of $\alpha$ and $\beta$ indicating strong atmospheric turbulence conditions, and $G[.]$ is the Meijer's G function as defined in [18, Eq. (9.301)].

For the SNR of the FSO receiver with HD and IM/DD, a unified expression can be given by $\gamma_r=(\eta_e I)^r/N_0$, where $\eta_e$ is the optical-to-electrical conversion ratio, $r$ is the parameter which denotes the type of detection being used (i.e. $r=1$ is associated with HD and $r=2$ associated with IM/DD), and $N_0$ is the additive white Gaussian noise. Therefore, a unified expression of PDF of $\gamma_r$ including both HD and IM/DD can be derived from Eq. (1) and is given by [19, Eq. (2)],

$$f_\gamma^{FSO}(\gamma_r) = \frac{\xi^2}{r\Gamma(\alpha)\Gamma(\beta)\gamma} G_{1,3}^{3,0}\left[\alpha\beta h\left(\frac{\gamma_r}{\mu_r}\right)^{\frac{1}{r}} \middle| \begin{matrix} \xi^2+1 \\ \xi^2, \alpha, \beta \end{matrix} \right], \tag{2}$$

where $h = \xi^2/(\xi^2+1)$, and $\mu_r = (\mathbb{E}[\eta_e I])^r/N_0$ refers to the average electrical SNR. In particular, for $r = 1$,

$$\mu_1 = \mu_{HD} = A_0 \eta_e \xi^2 / [(1+\xi^2)N_0] = \bar{\gamma}_1, \tag{3}$$

and for $r = 2$,

$$\mu_2 = \mu_{IM/DD} = \frac{A_0^2 \eta_e^2 \xi^4}{(1+\xi^2)^2 N_0} = \frac{\alpha\beta\xi^2(\xi^2+2)}{(\alpha+1)(\beta+1)(\xi^2+1)} \bar{\gamma}_2. \tag{4}$$

Using [18, Eq. (3.381.4)], the moment generating function (MGF) of $\gamma_r$ can be given by

$$M_\gamma^{FSO}(s) = \mathbb{E}\left[e^{s\gamma_r}\right] \tag{5}$$

$$= \int_0^\infty f_\gamma^{FSO}(\gamma_r) e^{s\gamma_r} d\gamma_r$$

$$= \frac{\xi^2}{r\Gamma(\alpha)\Gamma(\beta)} H_{2,3}^{3,1}\left[\alpha\beta h\left(\frac{-1}{\mu_r s}\right)^{\frac{1}{r}} \middle| \begin{matrix} (1,1/r),(\xi^2+1,1) \\ (\xi^2,1),(\alpha,1),(\beta,1) \end{matrix} \right],$$

where $H[.]$ is the Fox's H function defined in [20], and $\mathbb{E}[.]$ is the mathematical expectation. The cumulative density function (CDF) can be given by [19, Eq. (7)],

$$F_\gamma^{FSO}(\gamma_r) = 1 - \frac{\xi^2}{\Gamma(\alpha)\Gamma(\beta)} G_{2,4}^{4,0}\left[\alpha\beta h\left(\frac{\gamma_r}{\mu_r}\right)^{\frac{1}{r}} \middle| \begin{matrix} 1, \xi^2+1 \\ 0, \xi^2, \alpha, \beta \end{matrix} \right]. \tag{6}$$

Hence, the OP which is defined as the probability when the SNR falls below a predetermined protection threshold $\gamma_{th}$ can be directly obtained by replacing $\gamma_r$ with $\gamma_{th}$ in Eq. (6), that is, $P_{out}^{FSO} =$

$F_\gamma^{FSO}(\gamma_{th})$. Using the CDF expression, a unified expression of the average BER can be given by [19, Eq. (22)]

$$P_e = \frac{\delta}{2\Gamma(p)} \sum_{k=1}^{n} q_k^p \int_0^\infty \gamma^{p-1} e^{-q_k \gamma} F_\gamma(\gamma) d\gamma, \tag{7}$$

where $n$, $\delta$, $p$, and $q_k$ depend on detection types and modulation being assumed according to Table I. It is worth noting that this expression is general enough to be used for different modulation schemes and can also be applicable to RF channels. As a result, the average BER of the FSO link can be given by

$$P_e^{FSO} = \frac{n\delta}{2} - \frac{\delta \xi^2}{2\Gamma(p)\Gamma(\alpha)\Gamma(\beta)} \sum_{k=1}^{n} H_{3,4}^{4,1} \left[ \alpha\beta h (q_k \mu_r)^{-\frac{1}{r}} \middle| \begin{array}{l} (1-p,1/r)(1,1)(\xi^2+1,1) \\ (0,1)(\xi^2,1)(\alpha,1)(\beta,1) \end{array} \right]. \tag{8}$$

The derivation of Eq. (8) is given in Appendix A.

*B. RF subsystem*

For the RF subsystem, the $\kappa$-$\mu$ shadowed fading which includes several RF channel models such as Nakagami-$m$, Rayleigh, Rician, $\kappa$-$\mu$, and shadowed Rician fading distributions offers better and much more flexible representations of practical fading than Rayleigh and Rician fading used in [13-15]. The PDF of instantaneous SNR $\gamma$ of the $\kappa$-$\mu$ shadowed fading with non-negative real shape parameters $\kappa$, $\mu$ and $m$ is expressed as [16, Eq. (4)]

$$f_\gamma^{RF}(\gamma) = \frac{\mu^\mu m^m (1+\kappa)^\mu}{\Gamma(\mu)\bar{\gamma}(\mu\kappa+m)^m} \left(\frac{\gamma}{\bar{\gamma}}\right)^{\mu-1} e^{-\frac{\mu(1+\kappa)\gamma}{\bar{\gamma}}} {}_1F_1\left(m,\mu; \frac{\mu^2\kappa(1+\kappa)}{\mu\kappa+m} \frac{\gamma}{\bar{\gamma}}\right), \tag{9}$$

where ${}_1F_1(.)$ is the confluent hypergeometric function [20]. However, the confluent hypergeometric function leads to a very complicated form if not unresolvable. In [17], the $\kappa$-$\mu$ shadowed distribution is expressed as a mixture of squared Nakagami-$\hat{m}$ distributions when $\mu$ and $m$ take integer values. Thus, the PDF can be expressed in closed-form in terms of a finite number of elementary functions (powers and exponentials). It is demonstrated in [17] such restriction has little effect in practice when fitting field measurements to the $\kappa$-$\mu$ shadowed distribution, while being extremely convenient for computation. The PDF with integer $\mu$ and $m$ can be expressed as [17, Eq. (12)]

$$f_\gamma^{RF}(\gamma) = \sum_{i=0}^{M} C_i \frac{\gamma^{m_i-1}}{(m_i-1)!} \frac{1}{\Omega_i^{m_i}} e^{-\frac{\gamma}{\Omega_i}}, \tag{10}$$

where parameters $m_i$, $M$ and $\Omega_i$ are expressed in [17, Table I] in terms of the parameters of $\kappa$-$\mu$ shadowed distribution, namely, $\kappa$, $\mu$ and $m$. Using [18, Eq. (9.301)], the MGF of SNR can be written as

$$M_\gamma^{RF}(s) = \int_0^\infty f_\gamma^{RF}(\gamma) e^{s\gamma} d\gamma = \sum_{i=0}^{M} \frac{C_i}{(m_i-1)!} G_{1,1}^{1,1}\left(-s\Omega_i \middle| \begin{array}{c} 1-m_i \\ 0 \end{array}\right). \tag{11}$$

The CDF of $\kappa$-$\mu$ shadowed distribution can be expressed as [17, Eq. (13)]

$$F_\gamma^{RF}(\gamma) = 1 - \sum_{i=0}^{M} C_i e^{-\frac{\gamma}{\Omega_i}} \sum_{r=0}^{m_i-1} \frac{1}{r!} \left(\frac{\gamma}{\Omega_i}\right)^r. \tag{12}$$

Therefore, the OP of the RF link can be given by $P_{out}^{RF} = F_\gamma^{RF}(\gamma_{th})$. By using Eq. (12) and [18, Eq. (3.381.4)], the average BER of the RF link can be written as

$$P_e^{RF} = \frac{\delta}{2\Gamma(p)} \sum_{k=1}^{n} q_k^p \int_0^{\infty} \gamma^{p-1} e^{-q_k \gamma} F_\gamma^{RF}(\gamma) d\gamma \qquad (13)$$

$$= \frac{\delta n}{2} - \frac{\delta}{2\Gamma(p)} \sum_{k=1}^{n} \sum_{i=0}^{M} \sum_{r=0}^{m_i-1} \frac{C_i \Gamma(p+r)}{r!} \frac{1}{\Omega_i^r (q_k + 1/\Omega_i)^{p+r}},$$

where $n$, $\delta$, $p$, and $q_k$ for different modulation schemes are given in Table I.

*C. FSO/RF link with diversity combining*

In the hybrid FSO/RF link, diversity combining is implemented in the receiver to combine the FSO sub-link and the RF sub-link. Two popular diversity combining schemes, SC and MRC, are employed for the analysis of the hybrid FSO/RF link.

For the SC receiver, the received signal of each sub-system is combined such that the SNR is the maximum of the two sub-links [13], i.e.,

$$\gamma^{SC} = \max(\gamma^{FSO}, \gamma^{RF}). \qquad (14)$$

TABLE I
PARAMETERS FOR DIFFERENT MODULATIONS

| Modulation Scheme | $\delta$ | $p$ | $q_k$ | $n$ | Detection Type |
|---|---|---|---|---|---|
| OOK | 1 | 1/2 | 1/2 | 1 | FSO(IM/DD), RF |
| M-PSK | $\frac{2}{\max(\log_2 M, 2)}$ | 1/2 | $\sin^2\left(\frac{(2k-1)\pi}{M}\right)$ | $\max\left(\frac{M}{4}, 1\right)$ | FSO(HD), RF |
| M-QAM | $\frac{4}{\log_2 M}\left(1 - \frac{1}{\sqrt{M}}\right)$ | 1/2 | $\frac{3(2k-1)^2}{2(M-1)}$ | $\frac{\sqrt{M}}{2}$ | FSO(HD), RF |

Then the CDF of the SC can be given by

$$\begin{aligned}
F_\gamma^{SC}(\gamma) &= F_\gamma^{FSO}(\gamma) F_\gamma^{RF}(\gamma) \\
&= 1 - \frac{\xi^2}{\Gamma(\alpha)\Gamma(\beta)} G_{2,4}^{4,0}\left[\alpha\beta h\left(\frac{\gamma_r}{\mu_r}\right)^{\frac{1}{r}} \Bigg| \begin{array}{c} 1, \xi^2+1 \\ 0, \xi^2, \alpha, \beta \end{array}\right] - \sum_{i=0}^{M}\sum_{r=0}^{m_i-1} \frac{C_i}{r!} e^{-\frac{\gamma}{\Omega_i}}\left(\frac{\gamma}{\Omega_i}\right)^r \\
&\quad + \frac{\xi^2}{\Gamma(\alpha)\Gamma(\beta)} \sum_{i=0}^{M}\sum_{r=0}^{m_i-1} \frac{C_i}{r!} e^{-\frac{\gamma}{\Omega_i}}\left(\frac{\gamma}{\Omega_i}\right)^r G_{2,4}^{4,0}\left[\alpha\beta h\left(\frac{\gamma}{\mu_r}\right)^{\frac{1}{r}} \Bigg| \begin{array}{c} 1, \xi^2+1 \\ 0, \xi^2, \alpha, \beta \end{array}\right] \\
&= F_\gamma^{FSO}(\gamma) - \sum_{i=0}^{M}\sum_{\rho=0}^{m_i-1} \frac{C_i}{\rho!} e^{-\frac{\gamma}{\Omega_i}}\left(\frac{\gamma}{\Omega_i}\right)^\rho \\
&\quad + \frac{\xi^2}{\Gamma(\alpha)\Gamma(\beta)} \sum_{i=0}^{M}\sum_{\rho=0}^{m_i-1} \frac{C_i}{\rho!} e^{-\frac{\gamma}{\Omega_i}}\left(\frac{\gamma}{\Omega_i}\right)^\rho G_{2,4}^{4,0}\left[\alpha\beta h\left(\frac{\gamma}{\mu_r}\right)^{\frac{1}{r}} \Bigg| \begin{array}{c} 1, \xi^2+1 \\ 0, \xi^2, \alpha, \beta \end{array}\right].
\end{aligned} \qquad (15)$$

The OP of the SC receiver can be obtained by $P_{out}^{MRC} = F_\gamma^{MRC}(\gamma_{th})$. The average BER for OOK, M-PSK and M-QAM with the SC receiver can be given by

$$P_e^{SC} = \frac{n\delta}{2} - \frac{\delta\xi^2}{2\Gamma(p)\Gamma(\alpha)\Gamma(\beta)} \sum_{k=1}^{n} H_{3,4}^{4,1}\left[\alpha\beta h(q_k\mu_r)^{-\frac{1}{r}} \middle| \begin{array}{l}(1-p,1/r)(1,1)(\xi^2+1,1)\\(0,1)(\xi^2+1)(\alpha+1)(\beta+1)\end{array}\right] \quad (16)$$

$$-\frac{\delta}{2\Gamma(p)} \sum_{k=1}^{n}\sum_{i=0}^{M}\sum_{r=0}^{m_i-1} \frac{C_i\Gamma(p+r)}{r!} \frac{1}{\Omega_i^r(q_k+\Omega_i^{-1})^{p+r}}$$

$$+\frac{\delta}{2\Gamma(p)} \frac{\xi^2}{\Gamma(\alpha)\Gamma(\beta)} \sum_{k=1}^{n}\sum_{i=0}^{M}\sum_{r=0}^{m_i-1} \frac{q_k^p C_i}{r!\Omega_i^r}(q_k+\Omega_i^{-1})^{-p-r} \cdot$$

$$H_{3,4}^{4,1}\left[\alpha\beta h((q_k+\Omega_i^{-1})\mu_r)^{-1/r} \middle| \begin{array}{l}(1-p-r,1/r)(1,1)(\xi^2+1,1)\\(0,1)(\xi^2,1)(\alpha,1)(\beta,1)\end{array}\right].$$

The derivation of Eq. (16) is given in Appendix B.

For the MRC receiver of the hybrid FSO/RF link, the received signal of each sub-link is combined such that the SNR of the signal at the output of the receiver is the sum of the SNR of each sub-link [7, 13], i.e.,

$$\gamma^{MRC} = \gamma^{FSO} + \gamma^{RF}. \quad (17)$$

Therefore, the MGF of the SNR of the MRC receiver can be given by

$$M_\gamma^{MRC}(s) = \mathbb{E}[e^{s\gamma^{MRC}}] = \mathbb{E}[e^{s\gamma^{FSO}}]\mathbb{E}[e^{s\gamma^{RF}}] = M_\gamma^{FSO}(s)M_\gamma^{RF}(s). \quad (18)$$

The CDF can be found from the inverse Laplace transform of $M_\gamma^{MRC}(-s)/s$, which can be expressed as

$$F_\gamma^{MRC}(\gamma) = \mathcal{L}^{-1}\left(M_\gamma^{MRC}(-s)/s\right) \quad (19)$$

$$= \frac{\xi^2}{r\Gamma(\alpha)\Gamma(\beta)} \sum_{i=0}^{M} \frac{C_i}{(m_i-1)!} H_{1,1:1,1:2,3}^{0,1:1,1:3,1}\left[\begin{array}{c}\frac{\gamma}{\Omega_i}\\ \alpha\beta h(\gamma/\mu_r)^{1/r}\end{array} \middle| \begin{array}{l}-:(1,1):(1,\frac{1}{r})(\xi^2+1,1)\\(0,1,\frac{1}{r}):(m_i,1):(\xi^2,1)(\alpha,1)(\beta,1)\end{array}\right].$$

The derivation of Eq. (19) is given in Appendix C.

The OP of the MRC receiver can be obtained by $P_{out}^{MRC} = F_\gamma^{MRC}(\gamma_{th})$. Finally, the average BER of OOK, M-PSK and M-QAM for the MRC receiver can be given by

$$P_e^{MRC} = \quad (20)$$

$$\frac{\delta\xi^2}{2r\Gamma(\alpha)\Gamma(\beta)\Gamma(p)} \cdot$$

$$\sum_{k=1}^{n}\sum_{i=0}^{M} \frac{C_i}{(m_i-1)!} H_{1,1:1,1:2,3}^{0,1:1,1:3,1}\left[\begin{array}{c}\frac{1}{q_k\Omega_i}\\ \alpha\beta h\left(\frac{1}{\mu_r q_k}\right)^{\frac{1}{r}}\end{array} \middle| \begin{array}{l}(p,1,\frac{1}{r}):(1,1):(1,\frac{1}{r})(\xi^2+1,1)\\(0,1,\frac{1}{r}):(m_i,1):(\xi^2,1)(\alpha,1)(\beta,1)\end{array}\right].$$

The derivation of Eq. (20) is given in Appendix D.

## 3. Numerical results

In this section, numerical results are provided to investigate the performance of the hybrid FSO/RF link compared to the single FSO link and the single RF link. The analytical plots are obtained by using the derived formulas in the previous section, and Monte Carlo simulation results are obtained through MATLAB to prove the correctness of the theoretical results.

For the FSO link, the channel model is the Gamma-Gamma fading with the effects of atmosphere for weak ($\alpha = 2.902, \beta = 2.51$), moderate ($\alpha = 2.296, \beta = 1.822$), and strong ($\alpha = 2.064, \beta = 1.342$) turbulence [21]. The RF link is supposed to be subjected to Rician shadowed fading ($\kappa = 5, \mu = 1, m = 2$). The pointing error $\xi$ is set to 1. The analytical OPs of the single FSO link and the hybrid FSO/RF link are calculated successively, and the simulated OPs are also plotted in the same figure.

Figs. 2(a)-2(c) illustrate the OP of the FSO link and the hybrid FSO/RF link versus the average SNR of the FSO link under weak, moderate and strong turbulence, respectively. In Fig. 2, the average SNR of the RF link is set to 10 dB.

As can be seen from Fig. 2(a), the HD is better than the IM/DD. When the average SNR of the single FSO link is 20 dB, the OP of the single FSO link under HD is $2.51\times10^{-2}$ while that under IM/DD increases to $1.63\times10^{-1}$. The performance of the hybrid FSO/RF link can be improved along with the improvement of the FSO sub-link, so the OP of the hybrid FSO/RF link under HD is also better than that under IM/DD. For example, the OP of the hybrid FSO/RF link with the SC receiver under HD is $2.91\times10^{-3}$ while that under IM/DD increases to $1.89\times10^{-2}$. Furthermore, the OP of the hybrid FSO/RF link with the MRC receiver under HD is $1.40\times10^{-3}$ while that under IM/DD increases to $1.26\times10^{-2}$.

From Fig. 2(a), the hybrid FSO/RF link has better performance than the single FSO link. Moreover, the MRC receiver performs better than the SC receiver although its practical implement is more complex. For example, when the average SNR of the FSO link is 30 dB and the detection type is HD, the OP of the single FSO link is $2.53\times10^{-3}$. The OP of the hybrid FSO/RF link with the SC receiver is improved to $2.90\times10^{-4}$ while that with the MRC receiver is further improved to $1.42\times10^{-4}$.

The single FSO link and the hybrid FSO/RF link under moderate and strong atmospheric turbulence shown in Figs. 2(b) and 2(c) also reveal similar informations as Fig. 2(a). From Fig. 2, we can also observe that the OP deteriorates along with the degradation of atmospheric turbulence. For example, when the average SNR of the FSO link is 20 dB, the OPs under weak, moderate and strong atmospheric turbulence are $2.51\times10^{-2}$, $3.63\times10^{-2}$ and $5.26\times10^{-2}$ for the single FSO link under HD and $2.91\times10^{-3}$, $4.30\times10^{-3}$, and $6.09\times10^{-3}$ for the hybrid FSO/RF link with the SC receiver under HD, which illustrates the OP increases along with the degradation of the FSO channel.

It is worth noting that the results mentioned above are all under the pointing error $\xi = 1$, and the performance of the proposed hybrids FSO/RF link is also affected by pointing errors. Fig. 3 shows the OP of the hybrid FSO/RF system under moderate atmospheric turbulence with 5-dB average SNR of the RF link with different pointing errors ($\xi = 10.45, 2.15, 1.45, 1.15, 1, 0.65$). We can see a larger pointing error (smaller value of $\xi$) leads to worse system performance. Moreover, for a large variation in pointing errors from $\xi = 10.45$ to $\xi = 2.15$, there is no significant degradation in the OP performance. For instance, if $OP = 1\times10^{-3}$, there is a SNR loss of about 1 dB at $\xi = 2.15$ than that at $\xi = 10.45$.

As can be seen from Figs. 2 and 3, the simulated results agree well with the analytical results. This observation justifies the correctness of the derived formulas.

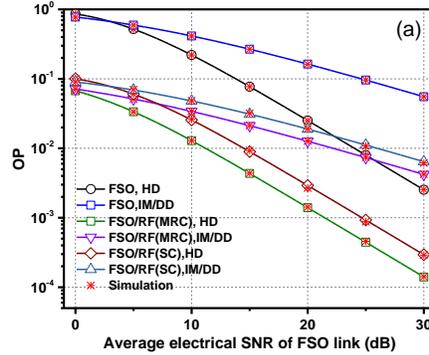

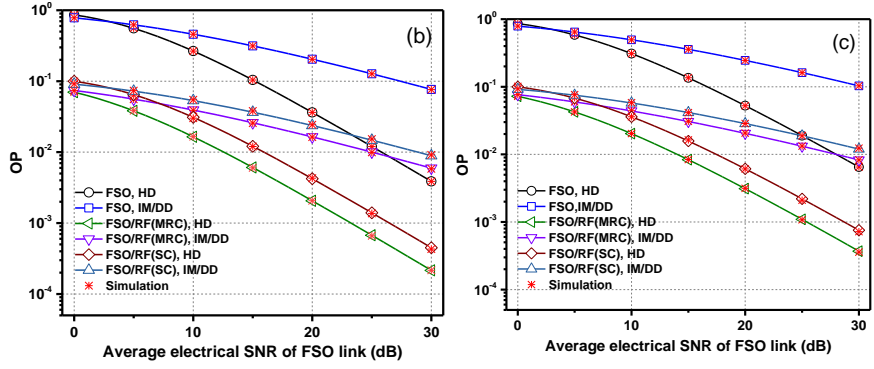

**Fig. 2.** OP of the single FSO link and the hybrid FSO/RF link versus average electrical SNR of the FSO link under HD and IM/DD with (a) weak, (b) moderate, (c) strong turbulent FSO channels.

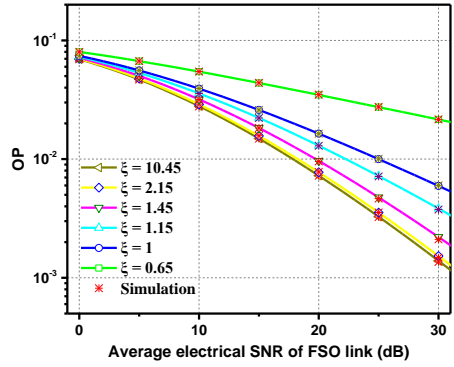

**Fig. 3.** OP of the hybrid FSO/RF link with varying effects of pointing errors.

Under the same RF link condition, the BERs for OOK, BPSK, QPSK, and 16-QAM of the single FSO link and the FSO/RF link are also investigated using the unified BER expressions, i.e. Eqs. (8), (16) and (20). The IM/DD is employed for OOK modulation while the HD is employed for BPSK, QPSK and 16-QAM modulation. Fig. 4 shows the average BERs for different modulation schemes of the FSO link and the hybrid FSO/RF link versus the average SNR of the FSO link under the pointing error $\xi = 1$. In each sub-figure of Fig. 4, the SNR of the RF link is set to 15 dB, and the BERs of the FSO system and the FSO/RF system under weak and strong atmospheric turbulence are plotted. A good matching between the simulated and analytical BER is evident from Fig. 4.

As can be seen from Fig. 4, the BER performance is greatly improved if an extra RF link is added to the FSO link and diversity combining is utilized. In our numerical investigation, when the average SNR of the FSO link is 20 dB, the BERs for OOK, BPSK, QPSK and 16-QAM of the FSO link under strong atmospheric turbulence are $7.48\times10^{-2}$, $7.05\times10^{-3}$, $1.29\times10^{-2}$, and $4.09\times10^{-2}$ while the average BERs for the hybrid FSO/RF link with the MRC receiver are reduced to $5.60\times10^{-3}$, $2.92\times10^{-4}$, $1.12\times10^{-3}$ and $1.29\times10^{-2}$, respectively.

From Fig. 4, we can also observe the BER of the MRC receiver is better than that of the SC receiver. For example, when the average SNR of the FSO link is 30 dB and the turbulence is weak, the BERs of the hybrid FSO/RF system with the SC receiver for OOK, BPSK, QPSK and 16-QAM are $1.85\times10^{-3}$, $2.78\times10^{-5}$, $1.15\times10^{-4}$, and $1.46\times10^{-3}$ while those with the MRC receiver are improved to $1.24\times10^{-3}$, $1.30\times10^{-5}$, $5.65\times10^{-5}$, and $8.95\times10^{-4}$, respectively.

Furthermore, we can see that as the effect of atmospheric turbulence decreases, the BER gets better. In Figs. 4(a)-(d), the BERs for OOK, BPSK, QPSK and 16-QAM of the hybrid FSO/RF system with the SC receiver are reduced from $7.73\times10^{-3}$, $2.40\times10^{-4}$, $2.11\times10^{-3}$, $1.97\times10^{-2}$ to $5.31\times10^{-3}$, $8.76\times10^{-5}$, $1.12\times10^{-3}$, $1.32\times10^{-2}$, respectively, when the atmospheric turbulence turns from strong to weak.

The pointing error on the BER performance of the hybrid FSO/RF link is also investigated. Fig. 5 shows the BER for 16-QAM of the hybrid FSO/RF link with 5-dB average SNR of the RF link under moderate atmospheric turbulence with different $\xi$ parameters ($\xi = 10.45, 2.15, 1.45, 1.15, 1, 0.65$). Similar to the OP, a larger pointing error also leads to worse BER performance. Moreover, when the $\xi$ characterizing the pointing error changes from a large value ($\xi = 10.45$) to a small value ($\xi = 2.15$) in our investigation, the BER performance does not degrade significantly. For example, at BER = $1\times10^{-3}$, there is a SNR loss less than 0.5 dB at $\xi = 2.15$ than that at $\xi = 10.45$.

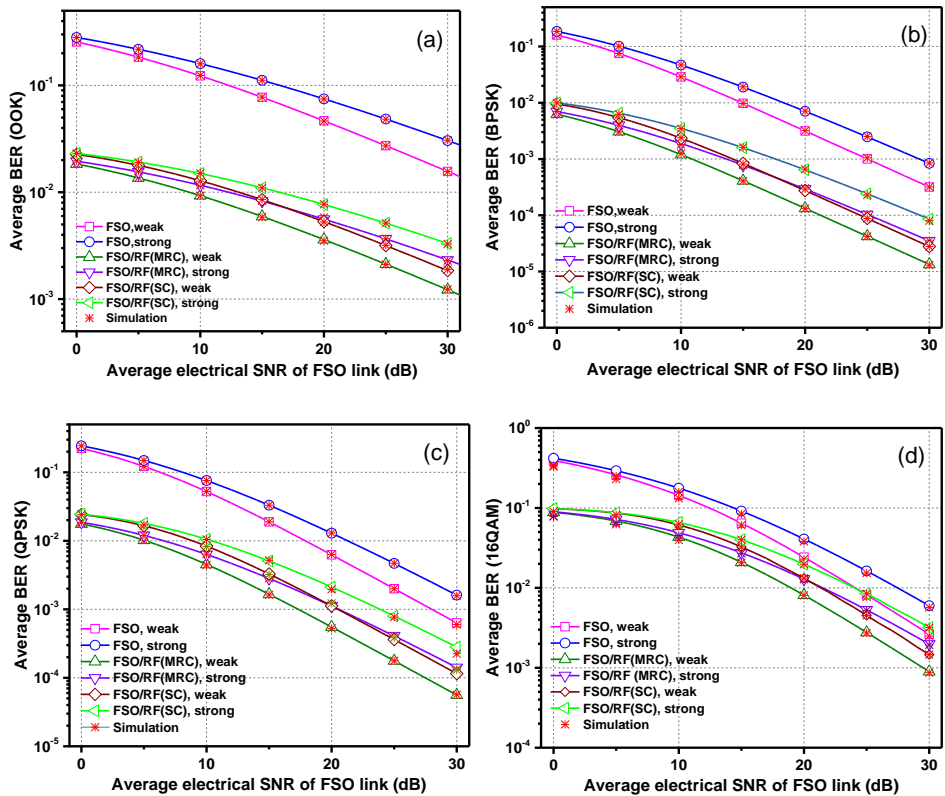

**Fig. 4.** Average BER for (a) OOK, (b) BPSK (c) QPSK, (d) 16QAM of the single FSO link and the hybrid FSO/RF link versus average electrical SNR of the FSO link with different average SNR of the RF link under weak and strong turbulence.

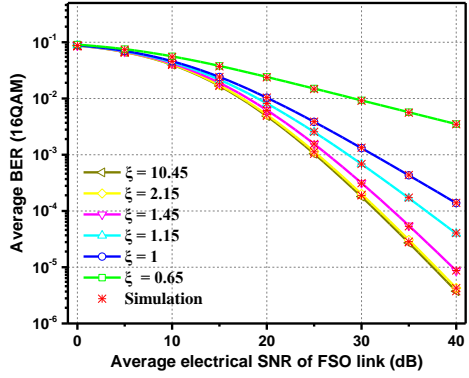

**Fig. 5. Average BER for 16-QAM of the hybrid FSO/RF link with varying effects of pointing errors.**

For the single RF link, the performance can also be improved if an extra FSO link is added and diversity combining is implemented. A more general RF fading channel which is $\kappa$-$\mu$ shadowed fading ($\kappa = 10$, $\mu = 2$ and $m = 1$) is considered in this case. For the FSO link, the atmospheric turbulence is set to moderate ($\alpha = 2.296$, $\beta = 1.822$), and the pointing error parameter $\xi$ is set to 1.

Fig. 6 shows the OP versus average SNR of the single RF link and the hybrid FSO/RF link with 5-dB and 15-dB average SNRs of the FSO link, respectively. We can see from Fig. 6 that the OP performance is improved if an extra FSO link is added to the RF link. For instance, when the average SNR of the RF link is 20 dB, the OP is $3.96\times10^{-3}$ while for the hybrid FSO/RF link under HD, the OP performance of the hybrid FSO/RF link with the MRC receiver is improved to $1.03\times10^{-3}$ and $1.53\times10^{-4}$ when the average SNRs of the FSO links are 5 dB and 15 dB, respectively.

Moreover, using the unified BER expressions given in Eq. (13), (16) and (20), the average BERs for different modulation schemes are presented. Figs. 7(a)-(d) show the BERs for OOK, BPSK, QPSK and 16-QAM of the single RF link and the hybrid FSO/RF link. In all cases, the average BER is reduced if an extra FSO link is enabled. According to our numerical investigation, the BERs for OOK, BPSK, QPSK, and 16-QAM of the single RF link are $1.25\times10^{-3}$, $3.63\times10^{-4}$, $1.25\times10^{-3}$ and $1.12\times10^{-2}$ of the single RF link while those of the hybrid FSO/RF link with the MRC receiver are reduced to $3.66\times10^{-4}$, $1.61\times10^{-5}$, $1.01\times10^{-4}$ and $2.86\times10^{-3}$, respectively, when the average SNR of the FSO link is 15 dB.

If a better FSO link is enabled, the performance of the hybrid FSO/RF link will be further improved. For example, when the average SNR of the single RF link is 20 dB, the BER for OOK is $1.25\times10^{-3}$. If the FSO link with 5-dB average SNR is enabled and the MRC receiver is used, the BER is improved to $6.46\times10^{-4}$. If the average SNR of the FSO link increases to 15 dB, the BER is further improved to $3.66\times10^{-4}$.

It can be observed from Figs. 6 and 7 that the simulation results match exactly to the derived analytical expressions obtained in this work.

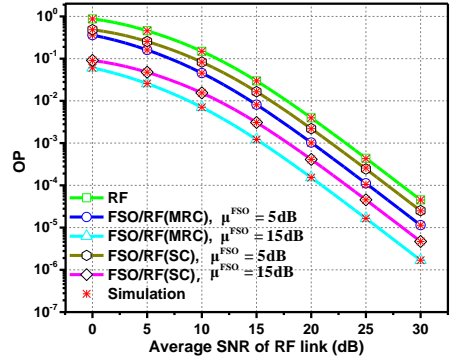

**Fig. 6.** OP of the single RF link and the hybrid FSO/RF link versus average SNR of the RF link with different average SNR of the FSO link.

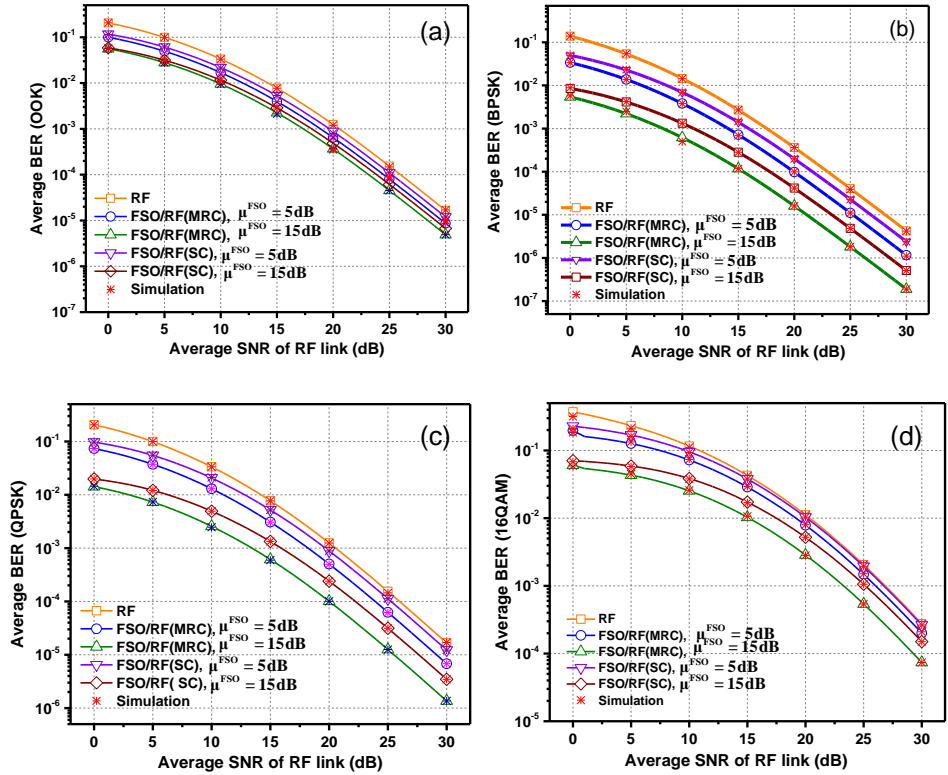

**Fig. 7.** Average BER for (a) OOK (b) BPSK (c) QPSK (d) 16-QAM of the single RF link and the hybrid FSO/RF link versus average SNR of the RF link with different average SNR of the FSO link.

## 4. Discussions

In summary, we analyzed the performance of hybrid FSO/RF systems with diversity combining which can be integrated in fiber-wireless MBH. We offered exact closed-form expressions of the OP and average BERs for different modulation schemes in terms of the Fox's H function. Since unified models of FSO and RF links are utilized, respectively, the derived expressions are very general, which accounts for atmospheric turbulence, pointing errors and two types of detection techniques (i.e. HD and IM/DD) for the FSO link and different RF fading models for the RF link.

Numerical results show the hybrid FSO/RF system with MRC has superior outage performance compared to single FSO systems and single RF systems in all atmospheric turbulence regimes, pointing errors and RF fading.

**Appendix**

*A. Derivation of average BER of FSO link*

In this appendix, we derive the average BER of the single FSO link that accounts for the HD and IM/DD as well as point errors. By substituting Eq. (6) to Eq. (7), the average BER of the single FSO link can be given by

$$\begin{aligned} P_e^{FSO} &= \frac{\delta}{2\Gamma(p)} \sum_{k=1}^{n} q_k^p \int_0^\infty \gamma_r^{p-1} e^{-q_k \gamma_r} F_\gamma^{FSO}(\gamma_r) d\gamma_r \\ &= \frac{\delta}{2\Gamma(p)} \sum_{k=1}^{n} q_k^p \int_0^\infty \gamma_r^{p-1} e^{-q_k \gamma_r} \left\{ 1 - \frac{\xi^2}{\Gamma(\alpha)\Gamma(\beta)} G_{2,4}^{4,0}\left[ \alpha\beta h\left(\frac{\gamma_r}{\mu_r}\right)^{\frac{1}{r}} \middle| \begin{matrix} 1, \xi^2+1 \\ 0, \xi^2, \alpha, \beta \end{matrix} \right] \right\} d\gamma_r \\ &= \frac{\delta}{2\Gamma(p)} \sum_{k=1}^{n} q_k^p (I_1 - I_2), \end{aligned} \quad (21)$$

where

$$I_1 = \int_0^\infty \gamma_r^{p-1} e^{-q_k \gamma} d\gamma_r = q_k^{-p} \Gamma(p), \quad (22)$$

by using [18, Eq. (3.381.4)], and

$$\begin{aligned} I_2 &= \int_0^\infty \gamma_r^{p-1} e^{-q_k \gamma_r} \frac{\xi^2}{\Gamma(\alpha)\Gamma(\beta)} G_{2,4}^{4,0}\left[ \alpha\beta h\left(\frac{\gamma_r}{\mu_r}\right)^{\frac{1}{r}} \middle| \begin{matrix} 1, \xi^2+1 \\ 0, \xi^2, \alpha, \beta \end{matrix} \right] d\gamma_r \\ &= \frac{\xi^2}{\Gamma(\alpha)\Gamma(\beta)} \cdot \int_0^\infty \gamma_r^{p-1} e^{-q_k \gamma_r} \cdot \frac{1}{2\pi j} \int_C \frac{\Gamma(-s)\Gamma(\xi^2-s)\Gamma(\alpha-s)\Gamma(\beta-s)}{\Gamma(1-s)\Gamma(\xi^2+1-s)} \left[ \alpha\beta h\left(\frac{\gamma_r}{\mu_r}\right)^{\frac{1}{r}} \right]^s ds \cdot d\gamma_r \\ &= \frac{\xi^2}{\Gamma(\alpha)\Gamma(\beta)} \cdot \frac{1}{2\pi j} \int_L \frac{\Gamma(-s)\Gamma(\xi^2-s)\Gamma(\alpha-s)\Gamma(\beta-s)}{\Gamma(1-s)\Gamma(\xi^2+1-s)} \left[ \alpha\beta h\left(\frac{1}{\mu_r}\right)^{\frac{1}{r}} \right]^s ds \int_0^\infty \gamma_r^{p-1+s/r} e^{-q_k \gamma_r} d\gamma_r \\ &= \frac{\xi^2}{\Gamma(\alpha)\Gamma(\beta)} \cdot \frac{1}{2\pi j} \int_L q_k^{-p-s/r} \Gamma(p+s/r) \frac{\Gamma(-s)\Gamma(\xi^2-s)\Gamma(\alpha-s)\Gamma(\beta-s)}{\Gamma(1-s)\Gamma(\xi^2+1-s)} \left[ \alpha\beta h\left(\frac{1}{\mu_r}\right)^{\frac{1}{r}} \right]^s ds \\ &= \frac{\xi^2}{\Gamma(\alpha)\Gamma(\beta)} \cdot q_k^{-p} H_{3,4}^{4,1}\left[ \alpha\beta h(q_k \mu_r)^{-\frac{1}{r}} \middle| \begin{matrix} (1-p, 1/r)(1,1)(\xi^2+1,1) \\ (0,1)(\xi^2,1)(\alpha,1)(\beta,1) \end{matrix} \right], \end{aligned} \quad (23)$$

by using [18, Eq. (3.381.4)] and [20].

Finally, by substituting Eqs. (22) and (23) to Eq. (21), Eq. (21) can be written as Eq. (8).

*B. Derivation of average BER of FSO/RF link with SC receiver*

Substituting Eq. (15) to Eq. (7), the average BER of the hybrid FSO/RF link with the SC receiver can be expressed as

$$P_e^{SC}$$
$$= \frac{\delta}{2\Gamma(p)} \sum_{k=1}^{n} q_k^p \int_0^{\infty} \gamma^{p-1} e^{-q_k \gamma} F_\gamma^{SC}(\gamma) d\gamma$$
$$=$$
$$\underbrace{\frac{\delta}{2\Gamma(p)} \sum_{k=1}^{n} q_k^p \int_0^{\infty} d\gamma \cdot \gamma^{p-1} e^{-q_k \gamma} \left\{ 1 - \frac{\xi^2}{\Gamma(\alpha)\Gamma(\beta)} G_{2,4}^{4,0}\left[ \alpha\beta h \left(\frac{\gamma_r}{\mu_r}\right)^{\frac{1}{r}} \middle| \begin{array}{c} 1, \xi^2+1 \\ 0, \xi^2, \alpha, \beta \end{array} \right] \right\}}_{P_1}$$
$$\underbrace{- \frac{\delta}{2\Gamma(p)} \sum_{k=1}^{n} q_k^p \int_0^{\infty} d\gamma \cdot \gamma^{p-1} e^{-q_k \gamma} \left[ \sum_{i=0}^{M} \sum_{r=0}^{m_i-1} \frac{C_i}{r!} e^{-\frac{\gamma}{\Omega_i}} \left(\frac{\gamma}{\Omega_i}\right)^r \right]}_{P_2}$$
$$\underbrace{+ \frac{\delta}{2\Gamma(p)} \sum_{k=1}^{n} q_k^p \int_0^{\infty} d\gamma \cdot \gamma^{p-1} e^{-q_k \gamma} \left\{ \frac{\xi^2}{\Gamma(\alpha)\Gamma(\beta)} \sum_{i=0}^{M} \sum_{r=0}^{m_i-1} \frac{C_i}{r!} e^{-\frac{\gamma}{\Omega_i}} \left(\frac{\gamma}{\Omega_i}\right)^r G_{2,4}^{4,0}\left[ \alpha\beta h \left(\frac{\gamma}{\mu_r}\right)^{\frac{1}{r}} \middle| \begin{array}{c} 1, \xi^2+1 \\ 0, \xi^2, \alpha, \beta \end{array} \right] \right\}}_{P_3},$$
(24)

where
$$P_1 = \frac{n\delta}{2} - \frac{\delta\xi^2}{2\Gamma(p)\Gamma(\alpha)\Gamma(\beta)} \sum_{k=1}^{n} H_{3,4}^{4,1}\left[ \alpha\beta h(q_k \mu_r)^{-\frac{1}{r}} \middle| \begin{array}{c} (1-p,1/r)(1,1)(\xi^2+1,1) \\ (0,1)(\xi^2,1)(\alpha,1)(\beta,1) \end{array} \right],$$
(25)

by using [18, Eq. (3.381.4)] and [20],
$$P_2 = \frac{\delta}{2\Gamma(p)} \sum_{k=1}^{n} \sum_{i=0}^{M} \sum_{r=0}^{m_i-1} \frac{C_i \Gamma(p+r)}{r!} \frac{1}{\Omega_i^r (q_k + \Omega_i^{-1})^{p+r}},$$
(26)

by using [18, Eq. (3.381.4)], and

$$P_3 =$$
(27)
$$\frac{\delta}{2\Gamma(p)} \frac{\xi^2}{\Gamma(\alpha)\Gamma(\beta)}$$
$$\sum_{k=1}^{n} \sum_{i=0}^{M} \sum_{\rho=0}^{m_i-1} \frac{q_k^p C_i}{r! \Omega_i^r \left(q_k + \Omega_i^{-1}\right)^{p+r}} H_{3,4}^{4,1}\left[ \alpha\beta h \left((q_k + \Omega_i^{-1})\mu_r\right)^{-1/r} \middle| \begin{array}{c} (1-p-r,1/r)(1,1)(\xi^2+1,1) \\ (0,1)(\xi^2,1)(\alpha,1)(\beta,1) \end{array} \right],$$

by using [18, Eq. (3.381.4)] and [20].

Finally, by substituting Eqs. (25)-(27) to Eq. (24), Eq. (24) can be written as Eq. (16).

### C. Derivation of CDF of hybrid FSO/RF system with MRC receiver

We apply inverse Laplace transform to $M_\gamma^{MRC}(-s)/s$, and use [20, Eq. (9.301)] and [22, Eq. (1.1)]. Then the CDF of the hybrid FSO/RF link can be written as

$$F_\gamma^{MRC}(\gamma)$$

$$= \frac{1}{2\pi j} \int_{\sigma-j\infty}^{\sigma+j\infty} \frac{M_\gamma^{MRC}(-s)}{s} e^{s\gamma} ds$$

$$= \frac{1}{2\pi j} \frac{\xi^2}{r\Gamma(\alpha)\Gamma(\beta)} \sum_{i=0}^{M} \frac{C_i}{(m_i-1)!} \cdot$$

$$\int_{\sigma+j\infty}^{\sigma+j\infty} G_{1,1}^{1,1}\left[s\Omega_i \Big|_0^{1-m_i}\right] H_{2,3}^{3,1}\left[\alpha\beta h(\mu_r s)^{-1/r} \Big|_{(\xi^2,1)(\alpha,1)(\beta,1)}^{(1,1/r)(\xi^2+1,1)}\right] s^{-1} e^{s\gamma} ds$$

$$= \frac{\xi^2}{r\Gamma(\alpha)\Gamma(\beta)} \sum_{i=0}^{M} \frac{C_i}{(m_i-1)!} \cdot \left(\frac{1}{2\pi j}\right)^2 \cdot$$

$$\iint_{L_1 L_2} \Gamma(-u)\Gamma(m_i+u)\Omega_i^u \frac{\Gamma(\xi^2+t)\Gamma(\alpha+t)\Gamma(\beta+t)\Gamma(-t/r)}{\Gamma(\xi^2+1+t)} \left(\frac{\mu_r^{\frac{1}{r}}}{\alpha\beta h}\right)^t dudt$$

$$\cdot \left(\frac{1}{2\pi j}\right) \int_{\sigma-j\infty}^{\sigma+j\infty} s^{u+t/r-1} e^{sr} ds. \qquad (28)$$

Using [18, Eq. (17.13.1)], Eq. (28) can be further written as

$$F_\gamma^{MRC}(\gamma) = \frac{1}{2\pi j} \int_{\sigma-j\infty}^{\sigma+j\infty} \frac{M_\gamma^{MRC}(-s)}{s} e^{s\gamma} ds$$

$$= \frac{1}{2\pi j} \frac{\xi^2}{r\Gamma(\alpha)\Gamma(\beta)} \sum_{i=0}^{M} \frac{C_i}{(m_i-1)!}$$

$$\int_{\sigma+j\infty}^{\sigma+j\infty} G_{1,1}^{1,1}\left[s\Omega_i \Big|_0^{1-m_i}\right] H_{2,3}^{3,1}\left[\alpha\beta h(\mu_r s)^{-1/r} \Big|_{(\xi^2,1)(\alpha,1)(\beta,1)}^{(1,1/r)(\xi^2+1,1)}\right] s^{-1} e^{s\gamma} ds$$

$$= \frac{\xi^2}{r\Gamma(\alpha)\Gamma(\beta)} \sum_{i=0}^{M} \frac{C_i}{(m_i-1)!} \cdot \left(\frac{1}{2\pi j}\right)^2$$

$$\iint_{L_1 L_2} \Gamma(-u)\Gamma(m_i+u)\Omega_i^u \frac{\Gamma(\xi^2+t)\Gamma(\alpha+t)\Gamma(\beta+t)\Gamma(-t/r)}{\Gamma(\xi^2+1+t)} \left(\frac{\mu_r^{\frac{1}{r}}}{\alpha\beta h}\right)^t dudt$$

$$\cdot \left(\frac{1}{2\pi j}\right) \int_{\sigma-j\infty}^{\sigma+j\infty} s^{u+t/r-1} e^{sr} ds, \qquad (29)$$

where $H[.\,,.]$ is the bivariate Fox's $H$ function which is defined in [22, A.1].

***D. Derivation of average BER of hybrid FSO/RF system with MRC receiver***

We substitute Eq. (19) to Eq. (7) and utilize [23, Eq. (1.11)], and then the average BER of the hybrid FSO/RF systems can be given by

$$P_e^{MRC}$$

$$= \frac{\delta}{2\Gamma(p)} \sum_{k=1}^{n} q_k^p \int_0^\infty \gamma^{p-1} e^{-q_k\gamma} F_\gamma^{MRC}(\gamma) d\gamma$$

$$= \frac{\delta}{2\Gamma(p)} \sum_{k=1}^{n} q_k^p \int_0^\infty \gamma^{p-1} e^{-q_k\gamma}$$

$$\cdot \left[\frac{\xi^2}{r\Gamma(\alpha)\Gamma(\beta)} \sum_{i=0}^{M} \frac{C_i}{(m_i-1)!} \left(\frac{1}{2\pi j}\right)^2 \iint_{L_1 L_2} \frac{\Gamma(u)\Gamma(m_i-u)\Gamma(\xi^2-t)\Gamma(\alpha-t)\Gamma(\beta-t)\Gamma(t/r)}{\Gamma(\xi^2+1-t)\Gamma(u+t/r+1)} \left(\frac{\gamma}{\Omega_i}\right)^u \left(\alpha\beta h\left(\frac{\gamma}{\mu_r}\right)^{\frac{1}{r}}\right)^t dudt\right] d\gamma$$

$$= \frac{\delta\xi^2}{2r\Gamma(p)\Gamma(\alpha)\Gamma(\beta)} \sum_{k=1}^{n} \sum_{i=0}^{M} \frac{q_k^p C_i}{(m_i-1)!} \left(\frac{1}{2\pi j}\right)^2$$

$$\cdot \iint_{L_1 L_2} \frac{\Gamma(u)\Gamma(m_i-u)\Gamma(\xi^2-t)\Gamma(\alpha-t)\Gamma(\beta-t)\Gamma(t/r)}{\Gamma(\xi^2+1-t)\Gamma(u+t/r+1)} \left(\frac{1}{\Omega_i}\right)^u \left(\alpha\beta h\left(\frac{1}{\mu_r}\right)^{\frac{1}{r}}\right)^t dudt \int_0^\infty \gamma^{p-1+u+t/r} e^{-q_k\gamma} d\gamma. \qquad (30)$$

By using [18, Eq. (3.381.4)], Eq. (30) can be finally written as

$$P_e^{MRC} = \frac{\delta \xi^2}{2r\Gamma(p)\Gamma(\alpha)\Gamma(\beta)} \sum_{k=1}^{n} \sum_{i=0}^{M} \frac{C_i}{(m_i-1)!} (\frac{1}{2\pi j})^2$$

$$\iint_{L_1 L_2} \frac{\Gamma(u)\Gamma(m_i-u)\Gamma(\xi^2-t)\Gamma(\alpha-t)\Gamma(\beta-t)\Gamma(t/r)\Gamma(p+u+t/r)}{\Gamma(\xi^2+1-t)\Gamma(u+t/r+1)\Gamma(1+u+t/r)} (q_k\Omega_i)^u \left[\alpha\beta h(\mu_r q_k)^{-1/r}\right]^t dudt$$

$$= \frac{\delta \xi^2}{2r\Gamma(p)\Gamma(\alpha)\Gamma(\beta)} \sum_{k=1}^{n} \sum_{i=0}^{M} \frac{C_i}{(m_i-1)!} H_{1,1:1,1:2,3}^{0,1:1,1:3,1} \left[ \begin{array}{c} (q_k\Omega_i)^{-1} \\ \alpha\beta h(\mu_r q_k)^{-1/r} \end{array} \middle| \begin{array}{c} (p,1,1/r):(1,1):(1,1/r)(\xi^2+1,1) \\ (0,1,1/r):(m_i,1):(\xi^2,1)(\alpha,1)(\beta,1) \end{array} \right]. \quad (31)$$